\documentclass[ajl]{emulateapj}

\usepackage{graphicx,epsfig,natbib,color}
\usepackage{lscape}
\usepackage{graphicx}
\usepackage{epsfig}
\usepackage{natbib}
\usepackage[colorlinks, citecolor=blue]{hyperref}

\newcommand{\apjsub} {{ApJ submitted}}

\slugcomment{Submitted to ApJ}

\begin{document}
\title{Imaging and Spectroscopic Diagnostics on the Formation of Two Magnetic Flux Ropes Revealed by \textit{SDO}/AIA and \textit{IRIS}}

\author{X. Cheng$^{1,2,3}$, M. D. Ding$^{1,3}$, \& C. Fang$^{1,3}$}

\affil{$^1$School of Astronomy and Space Science, Nanjing University, Nanjing 210093, China}\email{xincheng@nju.edu.cn}
\affil{$^2$Key Laboratory of Solar Activity, National Astronomical Observatories, Chinese Academy of Sciences, Beijing 100012, China}
\affil{$^3$Key Laboratory for Modern Astronomy and Astrophysics (Nanjing University), Ministry of Education, Nanjing 210093, China}

\begin{abstract}
Helical magnetic flux rope (MFR) is a fundamental structure of corona mass ejections (CMEs) and has been discovered recently to exist as a sigmoidal channel structure prior to its eruption in the extreme ultraviolet (EUV) high temperature passbands of the Atmospheric Imaging Assembly (AIA). However, when and where the MFR is built up are still elusive. In this paper, we investigate two MFRs (MFR1 and MFR2) in detail, whose eruptions produced two energetic solar flares and CMEs on 2014 April 18 and 2014 September 10, respectively. The AIA EUV images reveal that for a long time prior to their eruption, both MFR1 and MFR2 are under formation, which is probably through magnetic reconnection between two groups of sheared arcades driven by the shearing and converging flows in the photosphere near the polarity inversion line. At the footpoints of the MFR1, the \textit{Interface Region Imaging Spectrograph} Si IV, C II, and Mg II lines exhibit weak to moderate redshifts and a non-thermal broadening in the pre-flare phase. However, a relatively large blueshift and an extremely strong non-thermal broadening are found at the formation site of the MFR2. These spectral features consolidate the proposition that the reconnection plays an important role in the formation of MFRs. For the MFR1, the reconnection outflow may propagate along its legs, penetrating into the transition region and the chromosphere at the footpoints. For the MFR2, the reconnection probably takes place in the lower atmosphere and results in the strong blueshift and non-thermal broadening for the Mg II, C II, and Si IV lines.
\end{abstract}

\keywords{Sun: corona --- Sun: coronal mass ejections (CMEs) --- Sun: magnetic fields --- Sun: UV radiation}
Online-only material: animations, color figures

\section{Introduction}

Magnetic flux rope (MFR) is a current channel with magnetic field lines wrapping around its central axis by more than one turn. Such a coherently helical structure frequently erupts from the Sun, forming a coronal mass ejection (CME) in the lower corona \citep{cheng13_driver,liurui14,chen11_review}, and then propagating into the interplanetary space taking a form of magnetic cloud as indicated by features including the magnetic
field rotation, the drop of density and proton
temperature, and the low plasma beta in the in-situ data \citep{burlaga81,zhang07_icme,liuying14}. 

Previous observations have shown that MFRs probably exist in the corona prior to their eruption. Based on the different origins, MFRs can be divided into two categories. One type originates from long decayed active regions and appears as dark cavities in visible or extreme ultraviolet (EUV) passbands when viewed at the solar limb \citep{gibson06_apj,regnier11,mackay10}. The spinning motions \citep{wangym10,lixing12}, bright ring \citep{dove11}, and ``lagomorphic" structure of linear polarization in the cavities \citep{bak-steslicka13} have been disclosed and considered to be strong evidence of helical structures. This type of MFRs also displays as filament channels when seen at the solar disk, the counterpart of the cavities, in which the emission is much weaker than the background in EUV passbands but the magnetic field is well organized and probably includes dips \citep{low95_apj,gibson06_jgr}. Distinguishing from cavities, the other group of MFRs stems from mature active regions, manifesting themselves as forward or reversed sigmoids in soft X-ray (SXR) and EUV passbands \citep{canfield99,sterling00,fan04,kliem04}. A number of independent case studies have shown that the strongly twisted field lines and the sigmoidal emission pattern can be successfully reproduced in extrapolated non-linear force free structures \citep{guo10_filament,savcheva12a,suyingna12,cheng13_double,jiang14_nlfff,inoue13,amari14}. 

Recently, \citet{zhang12} and \citet{cheng13_driver} discovered another piece of observational evidence for the existence of MFRs in mature active regions, i.e., an elongated EUV channel-like structure in the high temperature passbands of the Atmospheric Imaging Assembly \citep[AIA;][]{lemen12}. The hot channel initially appears as a twisted sigmoidal structure and then evolves to a semicircular one in the early rise phase. Quickly, it drives the formation and acceleration of a CME with the aid of the flare reconnection \citep{cheng11_fluxrope,cheng13_driver,lileping13}. Further studies including the continuous evolution of the MFR from the inner to outer corona as a coherent structure \citep{cheng14_tracking}, the relationship between the hot channel and associated prominence \citep{cheng14_kink,kumar11,liting13,yang14}, and the finding of reduced temperature but enhanced emission measure at the lower part of filament-associated hot channel \citep{chenbin14} further support that the hot channel is actually the helical MFR itself.

\begin{figure*}
\center {\includegraphics[width=15cm]{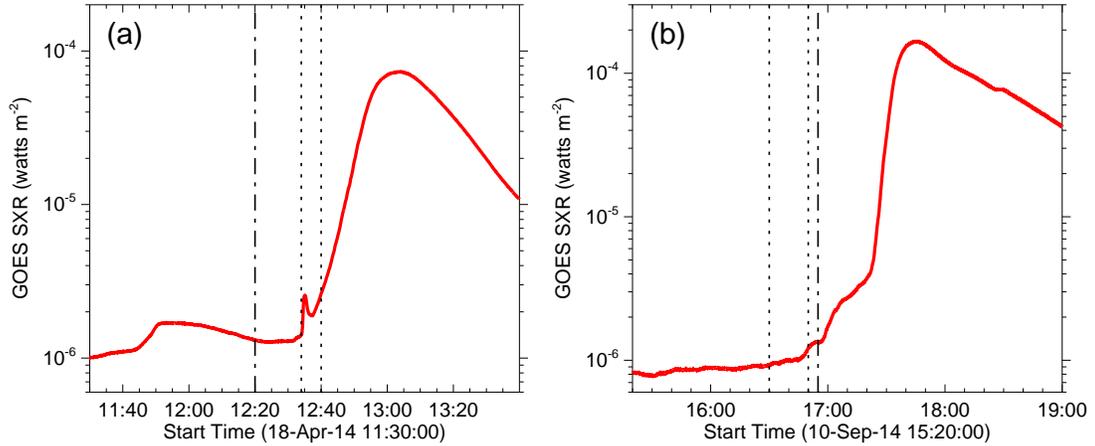}}
\caption{(a) \textit{GOES} soft X-ray 1--8~{\AA} flux of the solar flare (M7.3) on 2014 April 18. The vertical dash-dotted line denotes the instant at which the MFR1 is well formed as shown in Figure \ref{f1aia}; the two vertical dotted lines show the time interval, during which \textit{IRIS} observations are analyzed. (b) Same as in (a) but for the solar flare (X1.6) on 2014 September 10.}
\label{fgoes}
\end{figure*}

Although much evidence has revealed the existence of MFRs, when and where they are built up are still unknown. Theoretically, one possibility could be that the MFR is generated in the convection zone and partly emerges into the corona by buoyancy \citep[e.g.,][]{fan01,manchester04,archontis08_aa,leake13}. The other possibility is that the MFR is created directly in the corona either through flux cancellation in the photosphere prior to the eruption \citep{wangjx93,green09,green11,aulanier10,amari11,xia14,amari14} or by tether-cutting \citep{moore01,liur10}, breakout \citep{antiochos99,lynch08,karpen12}, and flare reconnection \citep{patsourakos13,kumar14,song14_formaion} in the corona during the eruption. In this paper, we investigate the formation and initiation of two MFRs and in particular, present for the first time the spectral features related to this process. The results suggest that the reconnection in the lower atmosphere may play an important role in forming and initiating the MFRs. The instruments and data reduction are introduced in Section 2. AIA Observations of the MFRs are shown in Section 3, followed by \textit{IRIS} observations of the MFRs in Section 4. Summary and discussions are given in Section 5.

\section{Instruments and Data Reduction}
The data sets are primarily from \textit{Solar Dynamics Observatory} \citep[\textit{SDO};][]{pesnell12} and newly launched Interface Region Imaging Spectrograph \citep[\textit{IRIS};][]{depontieu14}. The AIA on board \textit{SDO} images the solar atmosphere at temperatures ranging from 0.06 MK to 20 MK through 10 passbands (7 EUV passbands and 3 UV passbands). Each EUV (UV) image has a temporal cadence of 12 s (24 s) and spatial resolution of 1.2\arcsec. The Helioseismic and Magnetic Imager \citep[HMI;][]{schou12} also on board \textit{SDO} measures the vector magnetic field of the full solar disk. Here, we only take advantage of the line-of-sight magnetograms with the cadence of 720 s. Moreover, the \textit{GOES} provides the soft X-ray (SXR) 1--8 {\AA} flux of the whole Sun during the solar eruptions.

\textit{IRIS} obtains spectra and images from the photosphere, chromosphere, transition region, to corona with an unprecedented high spatial resolution of 0.33--0.4\arcsec, temporal cadence of $\sim$2 s, and spectral resolution of $\sim$1 km s$^{-1}$ over a field of view of 175\arcsec$\times$175\arcsec \citep{depontieu14}. The sit-and-stare mode is set during the periods of the formation and initiation of the two MFRs. The spectra include three doublet bright lines, Mg II h/k ($\sim$10$^{4}$ K) line forming in the chromosphere and C II 1334/1335 ($\sim$10$^{4.3}$ K) and Si IV 1394/1403 {\AA} ($\sim$10$^{4.8}$ K) lines in the transition region. In this study, we only analyze one component of the doublet lines, i.e., Mg II k 2796.347 {\AA}, C II 1335.7077 {\AA}, and Si IV 1402.77 {\AA}. The slit-jaw images (SJIs) with a cadence of 12 s in the 2796 {\AA}, 1400 {\AA}, and 1330 {\AA} passbands are taken simultaneously. All the spectra and images that we take for study are from calibrated level 2 data, which have already taken into account dark current subtraction, flat-field correction, and geometrical correction \citep{depontieu14,tian14_shock}. The wavelength calibration is critical for properly measuring the Doppler velocities. Here, the wavelengths of the C II and Si IV lines are calibrated by assuming a zero velocity of the nearby O I 1355.5977 {\AA} and Fe II 1405.608 {\AA} lines, respectively. For the Mg II line that is more perturbed by plasma dynamics, we calculate the average profile along the whole slit and find it can well represent the reference profile with an uncertainty of only 1--2 km s$^{-1}$. Moreover, we also eliminate the orbital variation of the line positions caused by the temperature change of the detector and the change of the distance between the spacecraft and the Sun with the routine $iris{\_}orbitval{\_}corr{\_}l2.pro$ \citep{tian15} in the SSW package.

\begin{figure*}
\center {\includegraphics[width=14cm]{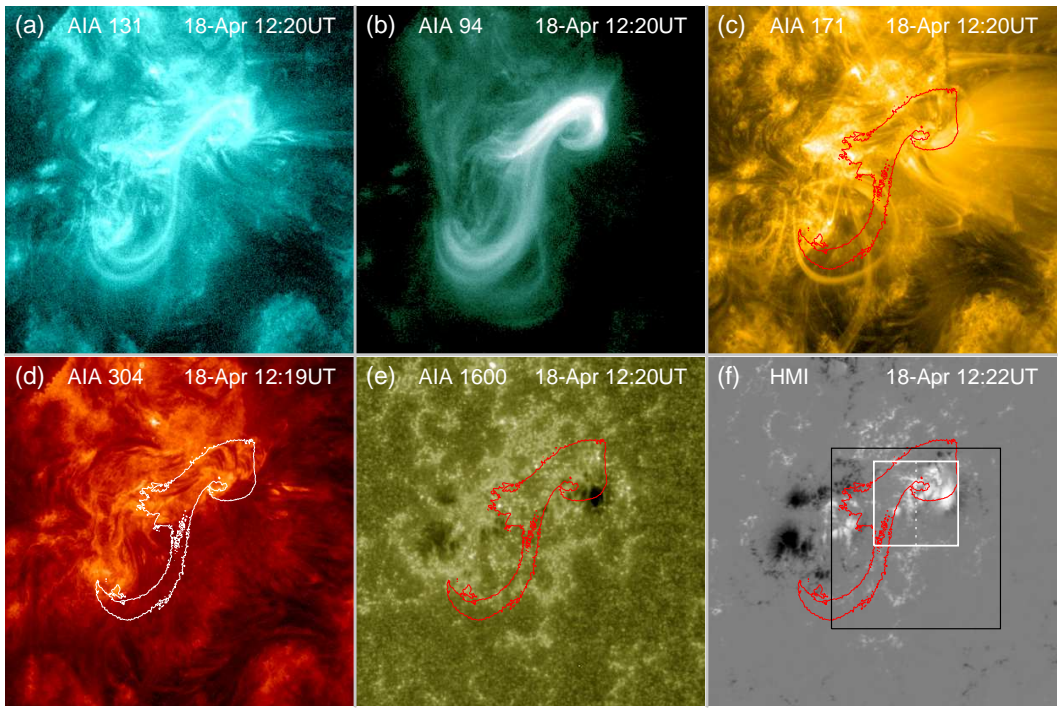}}
\caption{\textit{SDO}/AIA 131~{\AA} ($\sim$0.4, 10, and 16 MK), 94~{\AA} ($\sim$6 MK), 171~{\AA} ($\sim$0.6 MK), 1600~{\AA} ($\sim$0.1 MK), 1700~{\AA} ($\sim$0.5 MK) images, and \textit{SDO}/HMI line-of-sight magnetogram showing the MFR1 prior to the eruption (a and b), the overlying field (c), the lower atmosphere (d and e), and the magnetic field (f) in the active region NOAA 12036. The white and red contours in panels (c)--(f) indicate the location of the MFR1. The black box in panel (f) denotes the field-of-view of \textit{IRIS} (128\arcsec $\times$128\arcsec) and the smaller one (white; 60\arcsec $\times$60\arcsec) refers to the region which we are interesting in. The vertical dotted line indicates the position of \textit{IRIS} slit.}
(Animation of this figure is available in the online journal.)
\label{f1aia}
\end{figure*}

\section{AIA Observations of the MFRs}
In this section, we concentrate on EUV imaging of the formation of two MFRs (MFR1 and MFR2), whose eruptions produced a CME and an M7.3 flare on 2014 Aril 18 and a CME and an X1.6 flare on 2014 September 10, respectively. The GOES soft X-ray fluxes of the two flares are shown in Figure \ref{fgoes}.

\begin{figure*}
\center {\includegraphics[width=14cm]{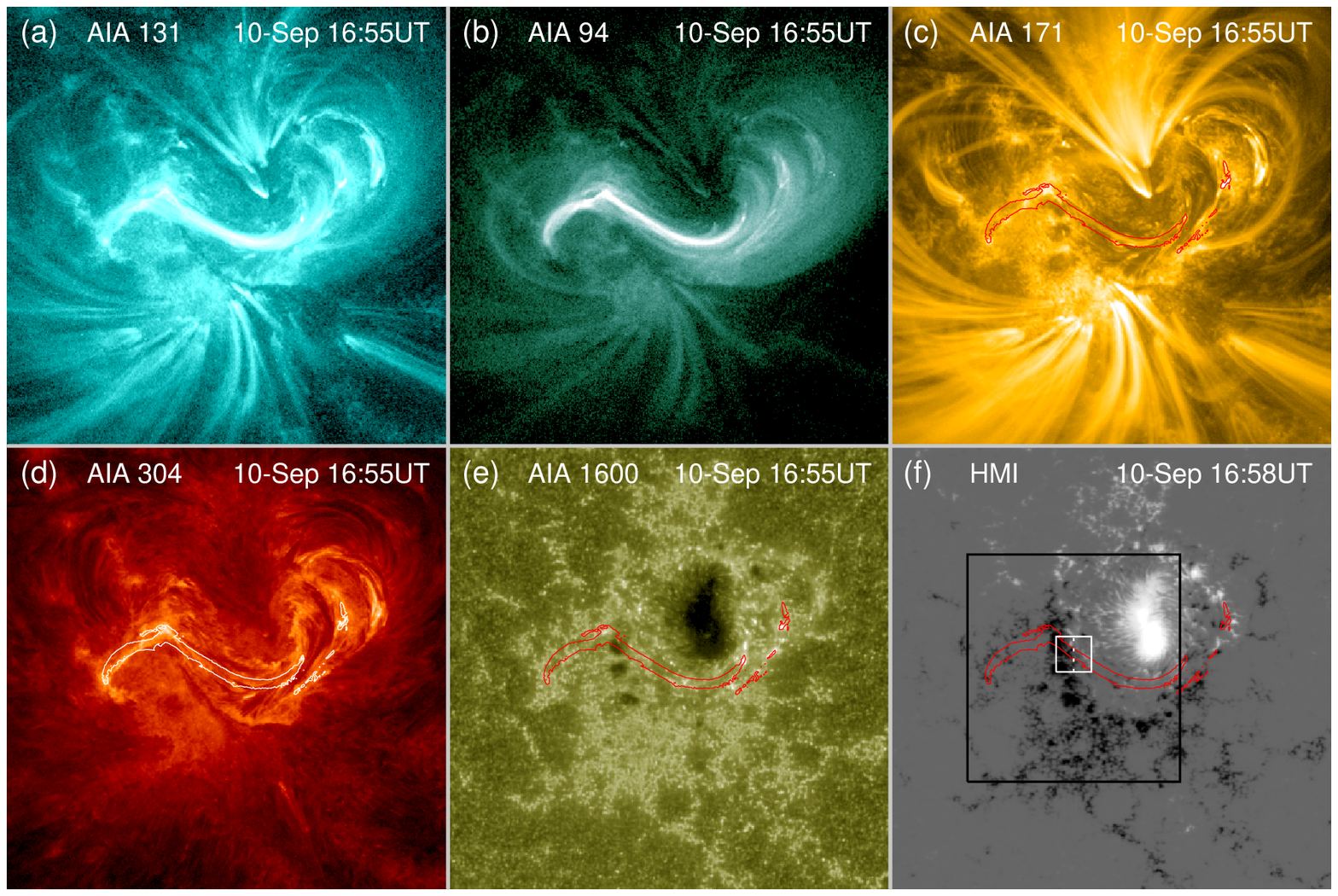}}
\caption{Same as Figure \ref{f1aia} but for the MFR2 in the active region NOAA 12158. The field of view of the black (white) box is 128\arcsec $\times$128\arcsec (20\arcsec $\times$20\arcsec).}
(Animation of this figure is available in the online journal.)
\label{f2aia}
\end{figure*}

\subsection{2014 April 18 MFR}

For the 2014 April 18 event, two groups of sheared arcades can be seen in the core field of associated active region at $\sim$07:30 UT. After $\sim$30 minutes, a confined flare (C4.8 class) begins and results in a quick brightening of the sheared arcades. In addition, some forward sigmoidal threads located above the arcades appear. It implies that part of the twisted field lines of MFR1 have probably been formed in the course of the confined flare \citep[also see;][]{patsourakos13}. After the flare, both the sheared arcades and sigmoidal threads disappear. At $\sim$11:20 UT, the sheared arcades appear again. After $\sim$15 minutes, some forward sigmoidal threads also show up. With the threads being further brightened, the concave middle part gradually becomes flat and detaches from the underlying arcades (for details please see online movie of Figure \ref{f1aia}). The whole evolution process is only visible in the AIA high temperature passbands (i.e., 131 {\AA} and 94 {\AA}) but not in other cooler passbands, indicating that the sheared arcades and sigmoidal threads are a very hot structure with a temperature of over 7 MK \citep{zhang12,cheng13_driver}. It suggests that the reconnection may exist continuously and help further build up the MFR1 after the confined flare.

The more interesting thing is that the right footpoints of the MFR1 start to become bright after $\sim$11:40 UT, which are visible in all AIA EUV and UV passbands. At the same time, some small loops underneath the central part of the MFR1 also brighten and appear as a cusp shape from the hot to cool temperature passbands in time sequence (see online movie of Figure \ref{f1aia}). The simultaneous brightening of the cusp-shaped small loops, the sigmoidal threads, and the footpoints of MFR1 seems to support that the reconnection could be of tether-cutting type, which takes place at the center of the sigmoid to convert the surrounding sheared arcades to the twisted lines of the MFR \citep[also see;][]{liur10,chenhd14}. At 12:20 UT, a well-shaped sigmoidal MFR1 has been formed (Figure \ref{f1aia}). After experiencing a slow rise and expansion phase, the MFR1 suddenly erupts and afterwards produces a strong flare and a CME. Note that, at $\sim$12:32 UT, a jet is triggered near the formation site of MFR1 and may play a role in initiating the impulsive acceleration of MFR1.

\subsection{2014 September 10 MFR}
For the 2014 September 10 event, during the long-term evolution before the eruption, the initial structure of the core field mainly consists of two groups of sheared arcades and central sigmoidal threads. Overall, these structures make up a reversed sigmoid. The sigmoid only appears in the AIA 94 {\AA} passband, showing that its temperature is very close to $\sim$7 MK, a peak temperature of the response function of this passband. Its being invisible in the AIA 131 {\AA} passband implies an absence of plasma with higher temperatures (e.g., 10 MK). At $\sim$13:57 UT, a C1.5 confined flare occurs near the right elbows of the sigmoid and make the nearby arcades brighten significantly, even in the AIA 131 {\AA} passband. After $\sim$15:00 UT, the most important evolution is repetitive appearance and disappearance of the central sigmoidal threads that could be a nascent structure of the MFR2 and have been developed via the reconnection in the course of the confined flare (for details please see online movie of Figure \ref{f2aia}). The sigmoidal threads initially appear at $\sim$15:10 UT. Subsequently, they become brighter and last for $\sim$20 minutes, and then disappear quickly. At $\sim$16:05 UT, the sigmoidal threads turn up again and last for only 10 minutes. At $\sim$16:30 UT, the most apparent feature is the brightening at the sigmoid center, probably the formation site of MFR2. After that ($\sim$16:45 UT), the sigmoidal threads get rapidly brightened and are also visible in the AIA 131 {\AA} passband. It implies that the sigmoidal threads have been heated to $\sim$10 MK at that time, most likely due to the reconnection. At $\sim$16:55 UT, a well-shaped sigmoidal structure, MFR2, has been formed as shown in Figure \ref{f2aia}. After experiencing a slow rise phase, it erupts and produces a strong flare and a CME. Similar to the MFR1, the intermittent appearance of the sigmoidal threads suggest the intermittent occurrence of reconnection, which not only heats the sigmoidal threads but also connects the sheared arcades to form the MFR2 prior to the eruption.

\section{\textit{IRIS} Observations of the MFRs}
\subsection{Transition Region and Chromospheric Imaging of the MFRs}

\begin{figure*}
\center {\includegraphics[width=14cm]{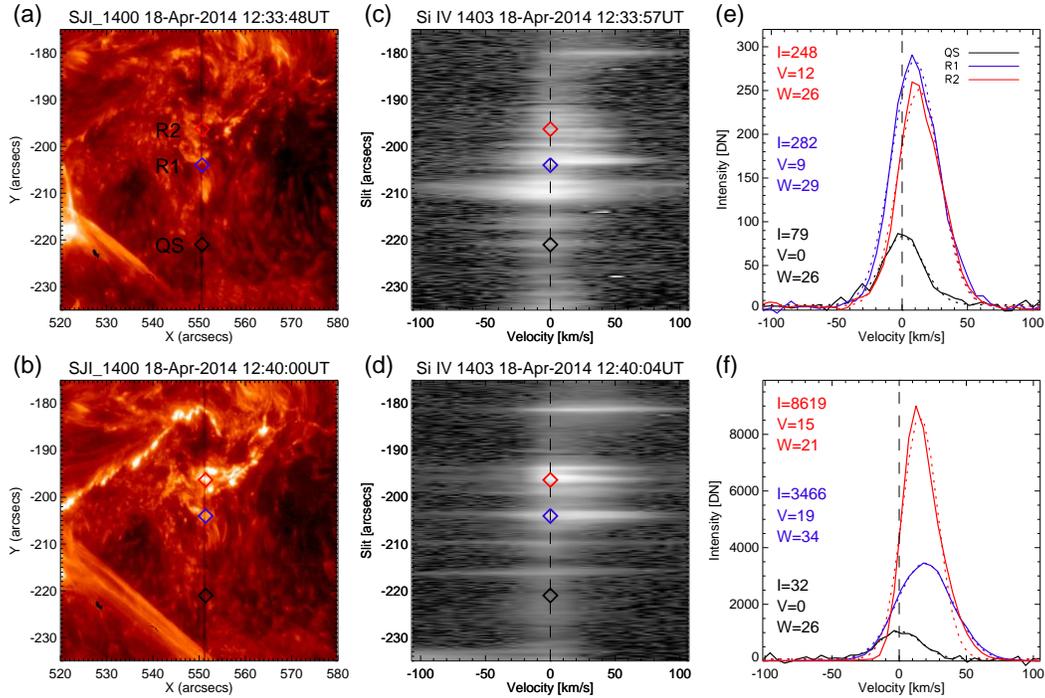}}
\caption{(a) and (b) \textit{IRIS} 1400 {\AA} SJI images displaying the footpoints of the MFR1 in the transition region. The blue (R1) and red (R2) diamonds denote the two pixels at the footpoints, and the black one (QS) indicates one pixel at the quiet region. (c) and (d) Spectra of the Si IV 1402.77 {\AA} line at the slit as shown in panels (a) and (b). (e) and (f) Profiles of the Si IV line at R1, R2, and QS. Their single Gaussian fittings are shown by the dotted curves. Quantities $I$, $V$, and $W$ correspond to the peak intensity, Doppler velocity, and FWHM. Note that, in panel f, the profile of the Si IV line at QS is multiplied by 30. Moreover, in panel c, the brightened area (y=[--213\arcsec, --207\arcsec]) in which the Si IV line is significantly blueshifted and broadened denotes the footpoints of a set of brightened coronal loops near the MFR1.}
(Animation of this figure is available in the online journal.)
\label{f1siiv}
\end{figure*}

\begin{figure*}
\center {\includegraphics[width=14cm]{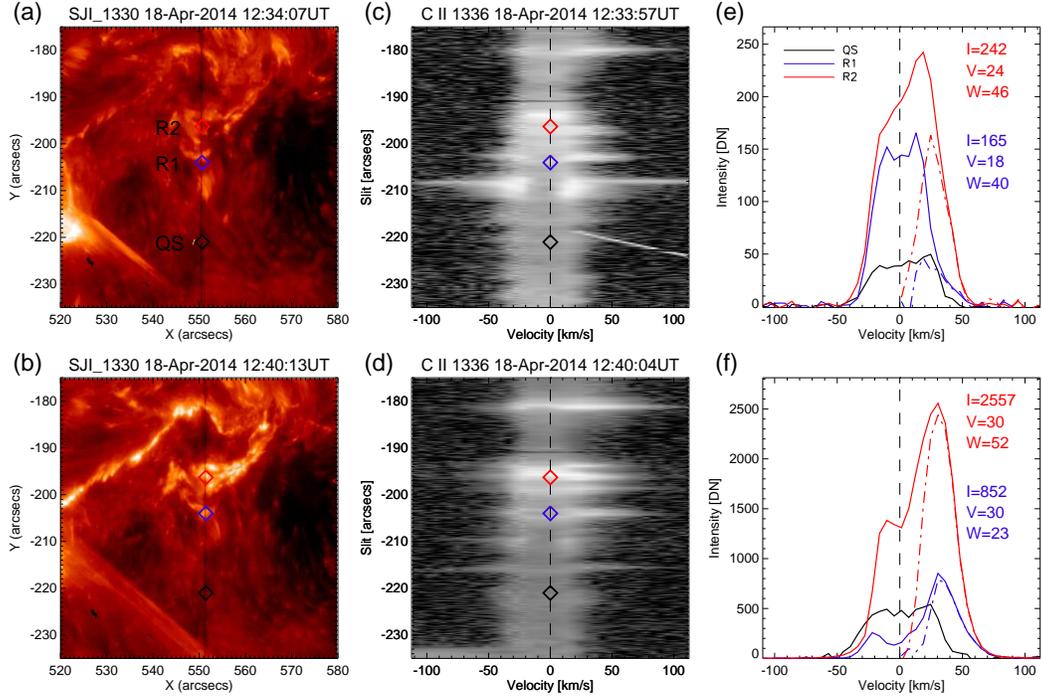}}
\caption{(a) and (b) \textit{IRIS} 1330 {\AA} SJI images displaying the footpoints of the MFR1 in the lower transition region. The red, blue, and black diamonds have the same meanings as in Figure \ref{f1siiv}. (c) and (d) Spectra of C II 1335.7077 {\AA} at the slit. (e) and (f) Profiles of C II 1335.7077 {\AA} at R1, R2, and QS. The dash-dotted curves denote the difference between the red wing and the blue wing. Note that, in panel f, the profile of the C II line at QS is multiplied by 10.}
\label{f1cii}
\end{figure*}

\begin{figure*}
\center {\includegraphics[width=14cm]{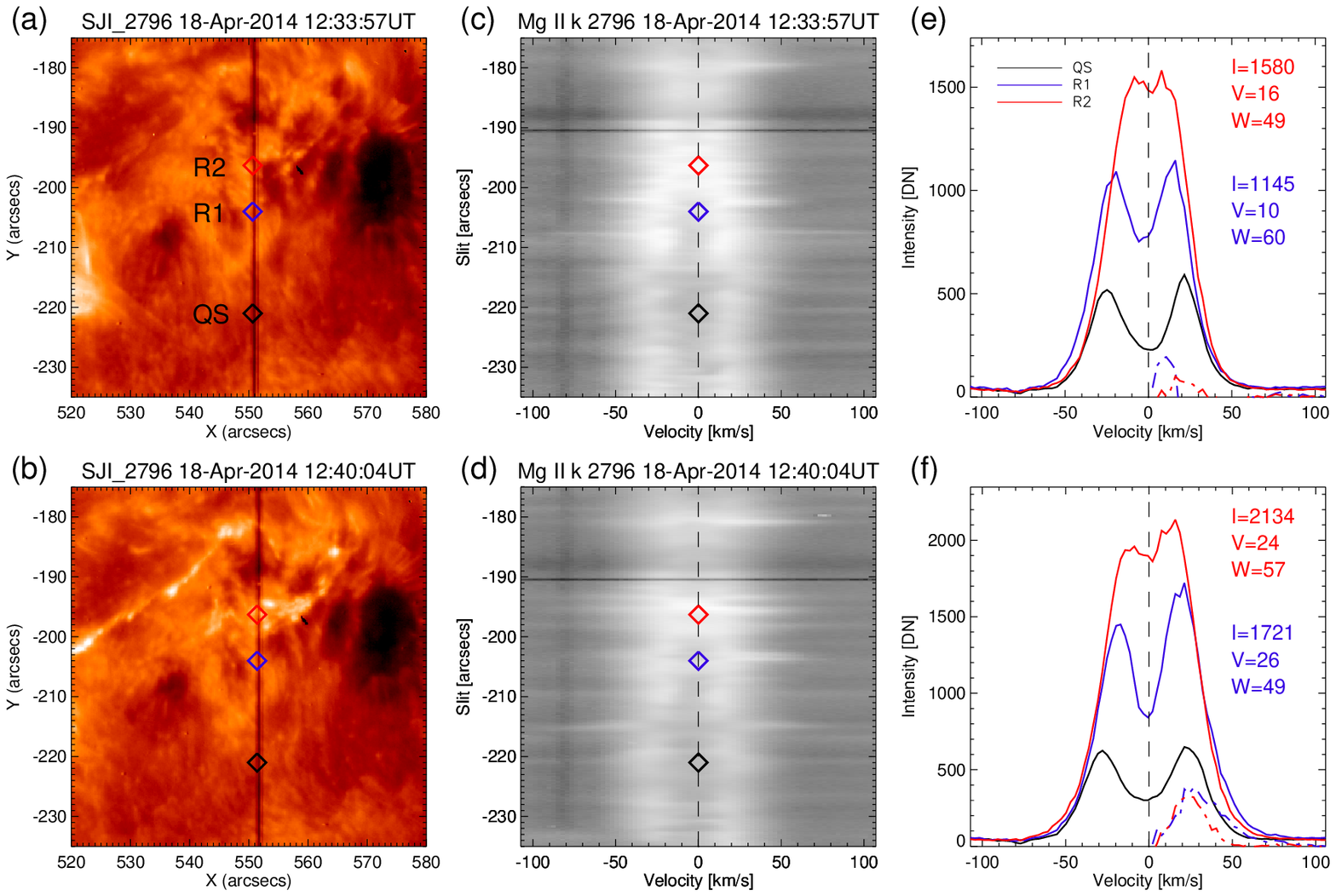}}
\caption{(a) and (b) \textit{IRIS} 2796 {\AA} SJI images displaying the footpoints of the MFR1 in the chromosphere. The red, blue, and black diamonds have the same meanings as in Figure \ref{f1siiv}. (c) and (d) Spectra of Mg II k 2796.347 {\AA} at the slit. (e) and (f) Profiles of Mg II k 2796.347 {\AA} at R1, R2, and QS. The dash-dotted curves denote the difference between the red wing and the blue wing.}
\label{f1mgii}
\end{figure*}

The pre-flare phase of the MFR1 and the formation phase of the MFR2 are also well observed by \textit{IRIS} with the sit-and-stare mode. For the MFR1, the slit is positioned crossing the right footpoints as shown in Figure \ref{f1aia}f. The whole observations are made from 12:33 UT to 17:18 UT. For the MFR2, the slit is positioned at the possible formation site as displayed in Figure \ref{f2aia}f. The observations last from 11:28 UT to 17:57 UT. Here, we mainly analyze the data in the intervals of 12:33--12:50 UT and 15:30--17:00 UT for the MFR1 and MFR2, respectively. From the \textit{IRIS} 1400 {\AA}, 1330 {\AA}, 2796 {\AA} images and the AIA 304 {\AA} and 1600 {\AA} images, we cannot see any signatures of the MFR1 (see the online movies of Figure \ref{f1aia} and Figure \ref{f1siiv}). It indicates that the magnetic structure of the MFR1 mainly lies in the corona with no stretching in the transition region and chromosphere except for the footpoints. The footpoints of the MFR1 start to brighten in the transition region and chromosphere at 12:35 UT. The snapshots of the footpoints that are strongly brightened at 12:40 UT are displayed in Figures \ref{f1siiv}b, \ref{f1cii}b, and \ref{f1mgii}b, as compared with the snapshots of the footpoints that are not yet strongly brightened in Figures \ref{f1siiv}a, \ref{f1cii}a, and \ref{f1mgii}a. 

\begin{figure*}
\center {\includegraphics[width=14cm]{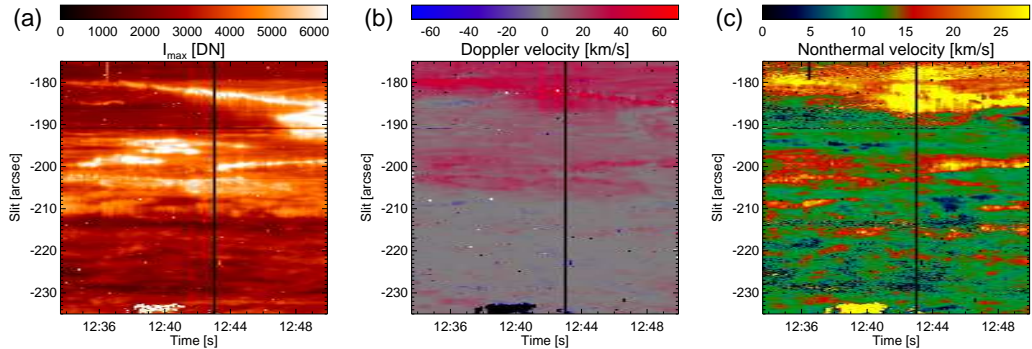}}
\caption{Slit-time plots of the peak intensity (a), Doppler velocity (b), and non-thermal velocity (c) of the Si IV 1402.77 {\AA} line for the 2014 April 18 event.}
\label{f1vel}
\end{figure*}

The appearance of the MFR2 is different from that of the MFR1. Besides in the 131 {\AA} and 94 {\AA} passbands, some bright sigmoidal loops are also present in the 1400 {\AA} and 2796 {\AA} passbands (Figures \ref{f2aia}, \ref{f2siiv}, \ref{f2mgii}, and attached movies). In the 171 {\AA} and 304 {\AA} passbands, some filamentary materials are also found to be cospatial with the threads in the middle part of the MFR2 (Figure \ref{f2aia}c and \ref{f2aia}d). Such a different emission feature implies that the height of the MFR2 may be lower than that of the MFR1, i.e., the MFR2 structure may partially extend into the transition region and even the chromosphere. It also suggests that  cool materials suspend in the dips of the MFR2 while it is not the case for the MFR1. The 1400 {\AA} and 2796 {\AA} SJI images of the brightened MFR2 at 16:30 UT are displayed in Figures \ref{f2siiv}a, \ref{f2cii}a, and \ref{f2mgii}a, while those after the brightening disappears ($\sim$16:50 UT) are shown in Figures \ref{f2siiv}b, \ref{f2cii}b, and \ref{f2mgii}b.

\subsection{Spectral Properties of the MFRs}

\subsubsection{Redshift at the Footpoints of the MFR1}
The spectra of \textit{IRIS} provide an important tool to diagnose the properties of the MFR in the lower atmosphere. Figures \ref{f1siiv}c, \ref{f1cii}c, and \ref{f1mgii}c and Figures \ref{f1siiv}d, \ref{f1cii}d, and \ref{f1mgii}d show the spectra of the Si IV 1402.77 {\AA}, C II 1335.7077 {\AA}, and Mg II k 2796.347 {\AA} lines at the slit before and after the footpoints are strongly brightened, respectively. Along the slit, we select three pixels: one corresponds to the quiet Sun region (QS in Figure \ref{f1siiv}a), and the other two are at the footpoints of the MFR1 (R1 and R2 in Figure \ref{f1siiv}a). The profiles of the Si IV line and the fitted ones at R1, R2, and QS are plotted in Figure \ref{f1siiv}e and \ref{f1siiv}f. One can see the following spectral features: (1) the profiles are nearly Gaussian; (2) the centroids of the line tend to be redshifted at R1 and R2; and (3) the peak intensities are significantly enhanced at R1 and R2 as compared with that at QS. In order to quantitatively measure the peak intensity, the Doppler shift, and the full width at half maximum (FWHM) of the Si IV line at the three pixels, we apply a single Gaussian fitting to the profiles. The results show that before the footpoints get brightened, the redshift velocities at R1 and R2 are 9 and 12 km s$^{-1}$, and the FWHMs are 29 and 26 km s$^{-1}$, respectively. With the footpoints being strongly brightened, the peak intensities increase as large as 10 times. The redshift velocities increase somewhat (19 and 15 km s$^{-1}$), but the FWHMs show little change (34 and 21 km s$^{-1}$). 

In order to study the temporal and spatial variation of the Si IV line, we apply a single Gaussian fitting to the profiles at the slit of interest (y=[--235\arcsec, --175\arcsec]) and obtain the slit-time plots of the peak intensity, the Doppler velocity, and the non-thermal velocity (Figure \ref{f1vel}). A bright ribbon exhibiting a very strong peak intensity and a relatively large Doppler velocity and non-thermal velocity can be identified. It actually denotes the flare ribbon sweeping across the slit. Besides the flare ribbon, one can see that the footpoints of the MFR1 also present a strong emission of the Si IV line. Moreover, such a strong emission is accompanied by a redshift of $\ge$15 km s$^{-1}$ and a non-thermal velocity of $\sim$15 km s$^{-1}$. These features start from 13:33 UT, or probably earlier, lasting for $\sim$13 minutes, and then gradually become weaker as the flare impulsive phase commences. Note that, the large Doppler velocity and non-thermal velocity at some pixels in the flare ribbon may be problematic because the Si IV line there exhibits a second component at the red wing, thus invalidating the single Gaussian assumption. 

The profiles of C II 1335.7077 {\AA} generally deviate from a single Gaussian shape. At QS (Figure \ref{f1cii}e and \ref{f1cii}f), the profile is rather flat at the central part with its centroid slightly shifted to the red wing. Compared with the QS, the C II line at the footpoints of the MFR1 shows a stronger emission and is possibly double-peaked. At R2, the red peak is much stronger. But at R1, the profile shows a weak central absorption with still a slightly stronger red peak (Figure \ref{f1cii}e). After the footpoints get strongly brightened, the emission is significantly enhanced especially at the red wing, which results in a more obvious line asymmetry. The intensities at the red peak increase by $\sim$10 times. Considering the significant line asymmetry and deviation from a single Gaussian, we adopt a simple method to derive the Doppler velocity. We subtract the intensity of the blue wing from the red one to get the residual emission whose peak is used to calculate the Doppler shift. Doing so is based on the assumption that the residual emission is from a moving plasma driven and heated by the reconnection in some place. In this way, the estimated Doppler velocities at R1 and R2 are $\sim$30 km s$^{-1}$ (Figure \ref{f1cii}f).

Owing to its chromospheric origin, the profiles of Mg II k 2796.347 {\AA} generally show a strong self-absorption in the line center due to a large opacity, in particular in the quiet region (e.g., at QS; Figure \ref{f1mgii}e and \ref{f1mgii}f). \cite{iris_leenaarts2,iris_leenaarts1} have studied in detail the diagnostic potential of this line at the quiet region based on radiation hydrodynamic simulations. Here, we present the characters of the line at the brightened region of the active region. At the footpoints of the MFR1, the line is enhanced with a reduced central absorption (k$_{3}$) at R1 or even no central absorption at R2 (Figure \ref{f1mgii}e). If also applying the method of subtraction of the blue wing (k$_{2v}$) from the red wing (k$_{2r}$), the redshift velocity is estimated to be $\sim$10 and 16 km s$^{-1}$ at R1 and R2, respectively. With the brightness at the footpoints of the MFR1 increasing, the redshift velocity becomes somewhat larger, i.e., $\sim$26 km s$^{-1}$ at R1 and $\sim$24 km s$^{-1}$ at R2 (Figure \ref{f1mgii}f).

\begin{figure*}
\center {\includegraphics[width=14cm]{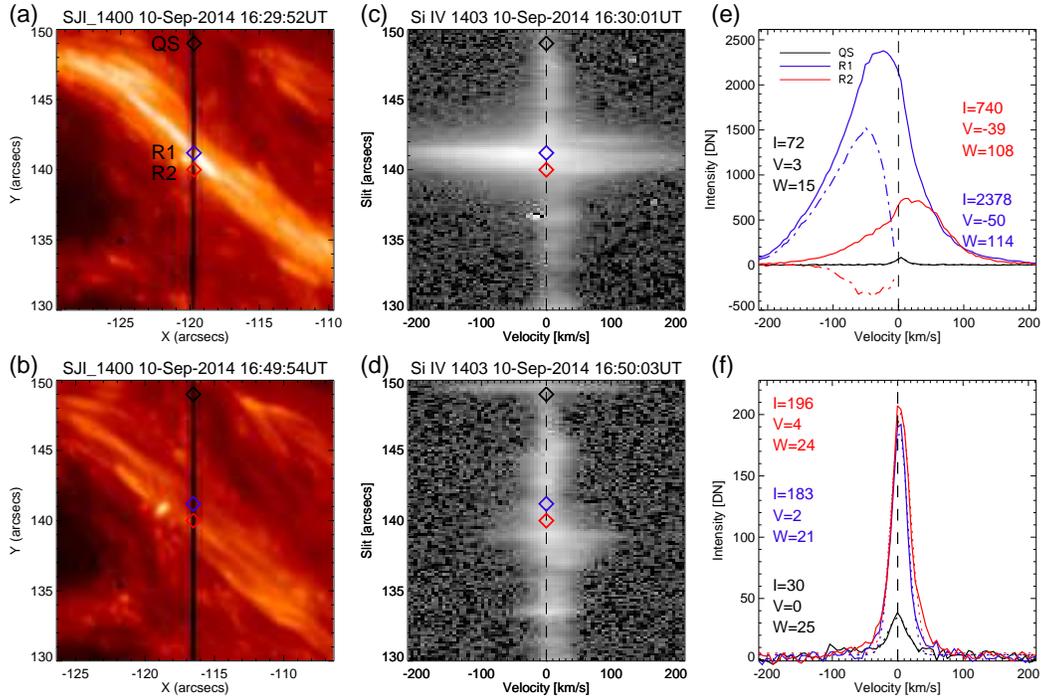}}
\caption{(a) and (b) \textit{IRIS} 1400 {\AA} SJI images displaying the center of the MFR2 in the transition region. The blue (R1) and red (R2) diamonds denote the two selected pixels at the MFR2 center, and the black one (QS) indicates one pixel at the quiet region. (c) and (d) Spectra of Si IV 1402.77 {\AA} at the slit. (e) and (f) Profiles of Si IV 1402.77 {\AA} at R1, R2, and QS. The profile of the blue wing subtracted by the corresponding red wing is shown as the dash-dotted curves.}
(Animation of this figure is available in the online journal.)
\label{f2siiv}
\end{figure*}

\subsubsection{Blueshift at the Center of the MFR2}
The Si IV 1402.77 {\AA} line at the formation site of the MFR2 shows a very strong Doppler blueshift and a large FWHM when it brightens, as shown in Figure \ref{f2siiv}c. However, when the brightness drops and returns to the background, the blueshift and the excess broadening of the line almost disappear, as shown in Figure \ref{f2siiv}d. We also select three pixels at the slit, R1 and R2 in the center of the MFR2 (the intensity at R1 is larger than that at R2), and QS in the quiet region. The profiles of the Si IV line at the three pixels are shown in Figure \ref{f2siiv}e and \ref{f2siiv}f. One can clearly see that all the three profiles are singly peaked in spite of quite different peak intensities (Figure \ref{f2siiv}f). The interesting fact is that when the line intensity increases, the line profile becomes broadened and asymmetric (Figure \ref{f2siiv}e). More specifically, the profile at R1 shows a stronger blue wing while that at R2 shows a stronger red wing. If we still use the single Gaussian fitting, the blueshift velocity is estimated to be $\sim$34 km s$^{-1}$ for R1 while the redshift velocity is $\sim$18 km s$^{-1}$ for R2. 

\begin{figure*}
\center {\includegraphics[width=14cm]{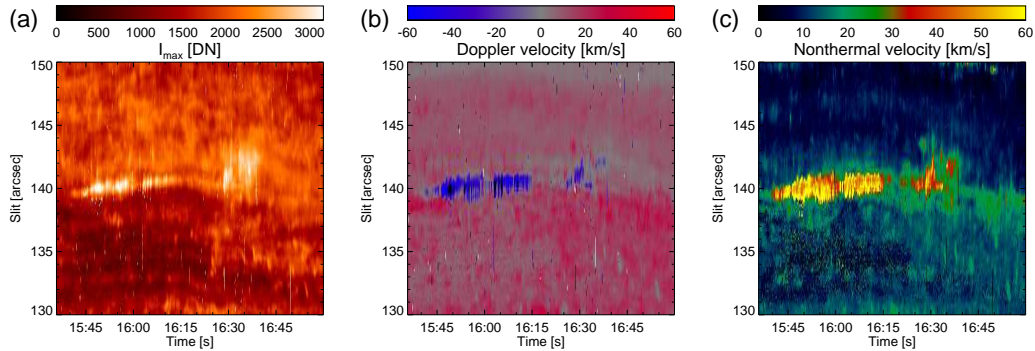}}
\caption{Slit-time plots of the peak intensity (a), Doppler velocity (b), and non-thermal velocity of the Si IV 1402.77 {\AA} line at the slit for the 2014 September 10 event.}
\label{f2vel}
\end{figure*}

\begin{figure*}
\center {\includegraphics[width=14cm]{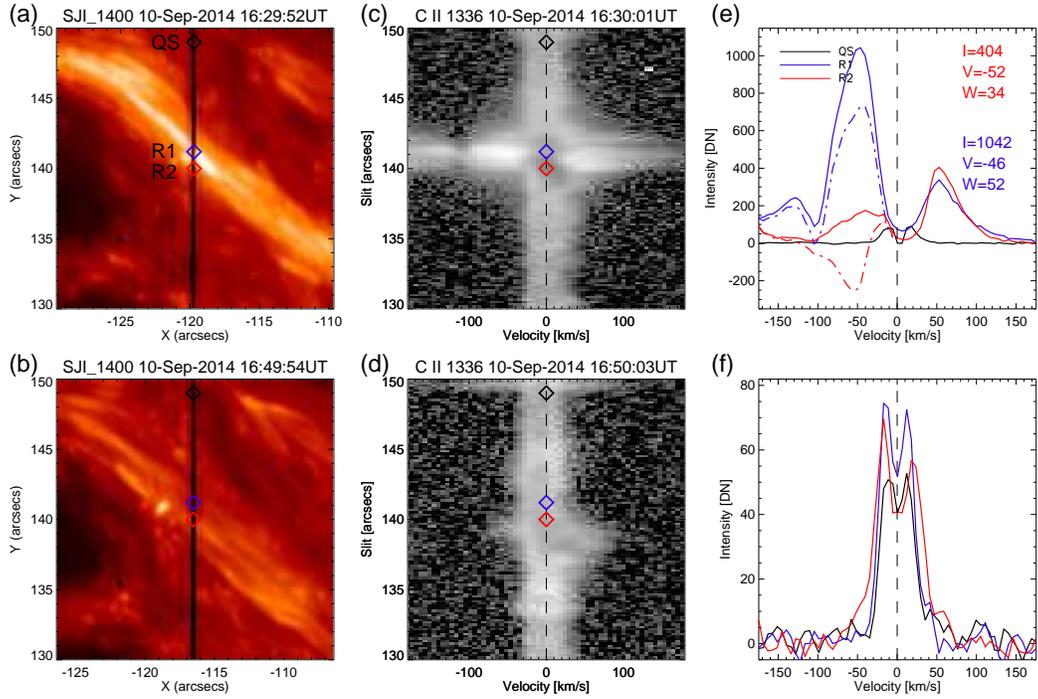}}
\caption{(a) and (b) \textit{IRIS} 1400 {\AA} SJI images displaying the center of the MFR2 in the transition region. The red, blue, and black diamonds have the same meanings as in Figure \ref{f2siiv}. (c) and (d) Spectra of C II 1335.7077 {\AA}. (e) and (f) Profiles of C II 1335.7077 {\AA} at R1, R2, and QS. The dash-dotted curves show the difference between the blue wing and the red wing.}
\label{f2cii}
\end{figure*}

\begin{figure*}
\center {\includegraphics[width=14cm]{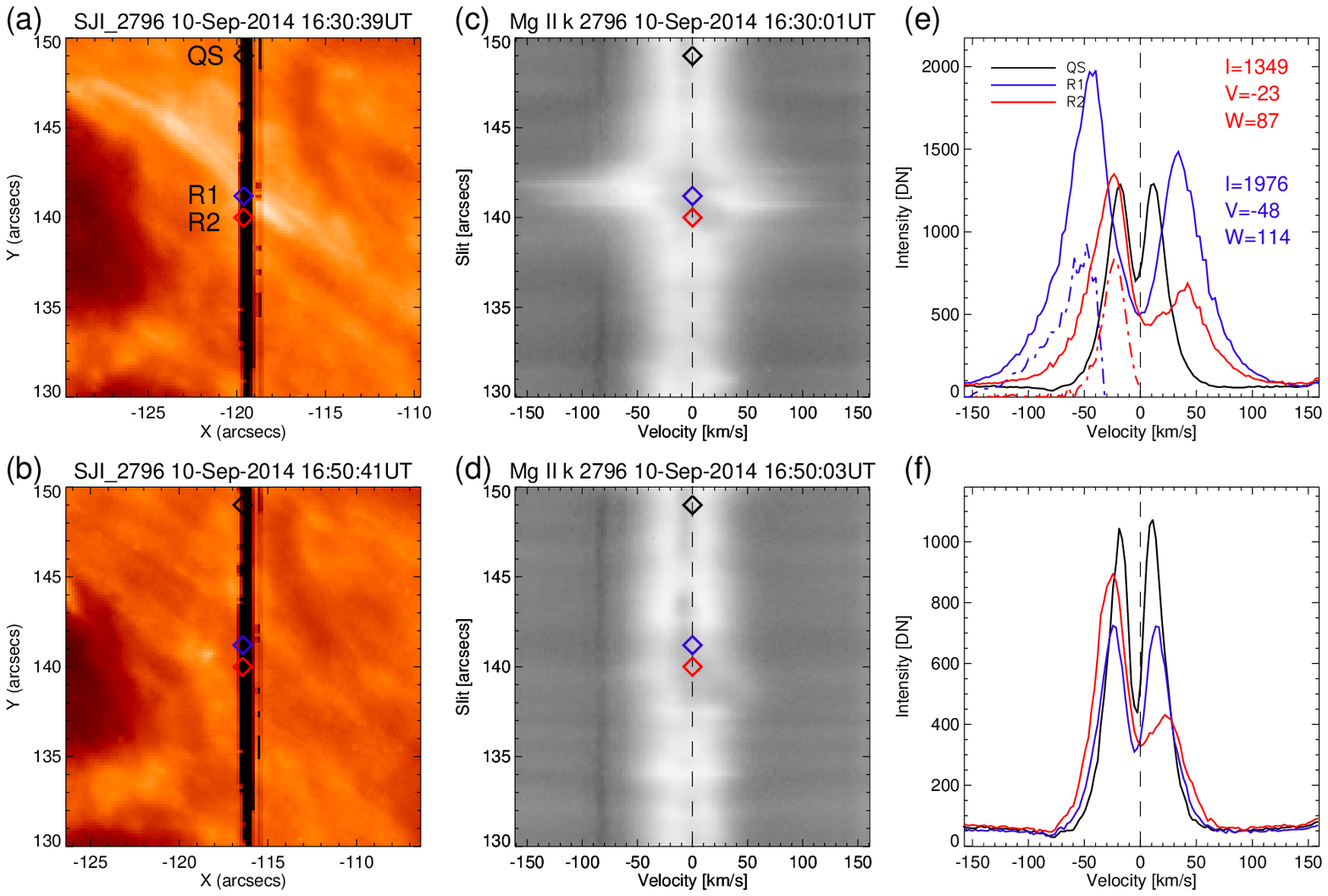}}
\caption{(a) and (b) \textit{IRIS} 2796 {\AA} SJI images displaying the MFR2 center in the chromosphere. The red, blue, and black diamonds have the same meanings as in Figure \ref{f2siiv}. (c) and (d) Spectra of Mg II k 2796.347 {\AA}. (e) and (f) Profiles of Mg II k 2796.347 {\AA} at R1, R2, and QS. The dash-dotted curves denote the profile of the blue wing subtracted by the red wing.}
\label{f2mgii}
\end{figure*}

The slit-time plots of the peak intensity, Doppler velocity, and non-thermal velocity at the slit of interest (y=[130\arcsec, 150\arcsec]) are plotted in Figure \ref{f2vel}. General speaking, the area with an enhanced brightness shows a strong Doppler blueshift velocity ($\ge$20 km s$^{-1}$) and non-thermal velocity ($\ge$40 km s$^{-1}$). Examining the profiles at R2, we find that they significantly deviate from a Gaussian shape especially at the blue wing. Considering the fact that the Mg II k line is blueshifted (Figure \ref{f2mgii}e and \ref{f2mgii}f), we suspect that the redshift at R2 could be due to an absorption at the blue wing. If this is the case, such an absorption could be caused by an upward moving plasma ejected from the chromosphere that is somewhat cooler and still absorptive relative to the background Si IV line emission. If adopting the method of the blue wing subtracted by the red one, the blueshift velocity is estimated to be $\sim$39 km s$^{-1}$ at R2 from an excess absorption and $\sim$50 km s$^{-1}$ at R1 from an excess emission. Note that not only the absolute values become significantly larger, but also the velocity sign is now different at R2, as compared with the results from the single Gaussian fitting.

Figure \ref{f2cii}c and \ref{f2cii}d present the spectra along the slit for C II 1335.7077 {\AA}. The profile of this line shows an extremely strong self-absorption in the line center and a complicated shape at line wings when the central part of MFR2 brightens (Figure \ref{f2cii}e). At R1, the blue wing is greatly enhanced, with emission even extending to --100 km s$^{-1}$. The blueshift velocity deduced from the blue wing subtracted by the red wing is $\sim$46 km s$^{-1}$. While at R2, the blue wing is significantly reduced relative to that in the red wing, as in the case of the Si IV line. If we still ascribe it to an absorption, the absorbing plasma should have a velocity of $\sim$52 km s$^{-1}$.

For the Mg II k 2796.347 {\AA} line, when it gets brightened, it is highly broadened with a stronger central absorption, especially at R1 (Figure \ref{f2mgii}c and \ref{f2mgii}e). Meanwhile, the blue peak (k$_{2v}$) is much stronger than the red peak (k$_{2r}$). We notice that the absorption center (k$_{3}$) does not change obviously. Therefore, we ascribe the stronger blue wing as being caused by a blueshifted emission. Using the difference profile between the blue wing and the red one, the blueshift velocity is estimated to be 48 km s$^{-1}$ at R1 and 23 km s$^{-1}$ at R2.
 
\begin{figure*}
\center {\includegraphics[width=14cm]{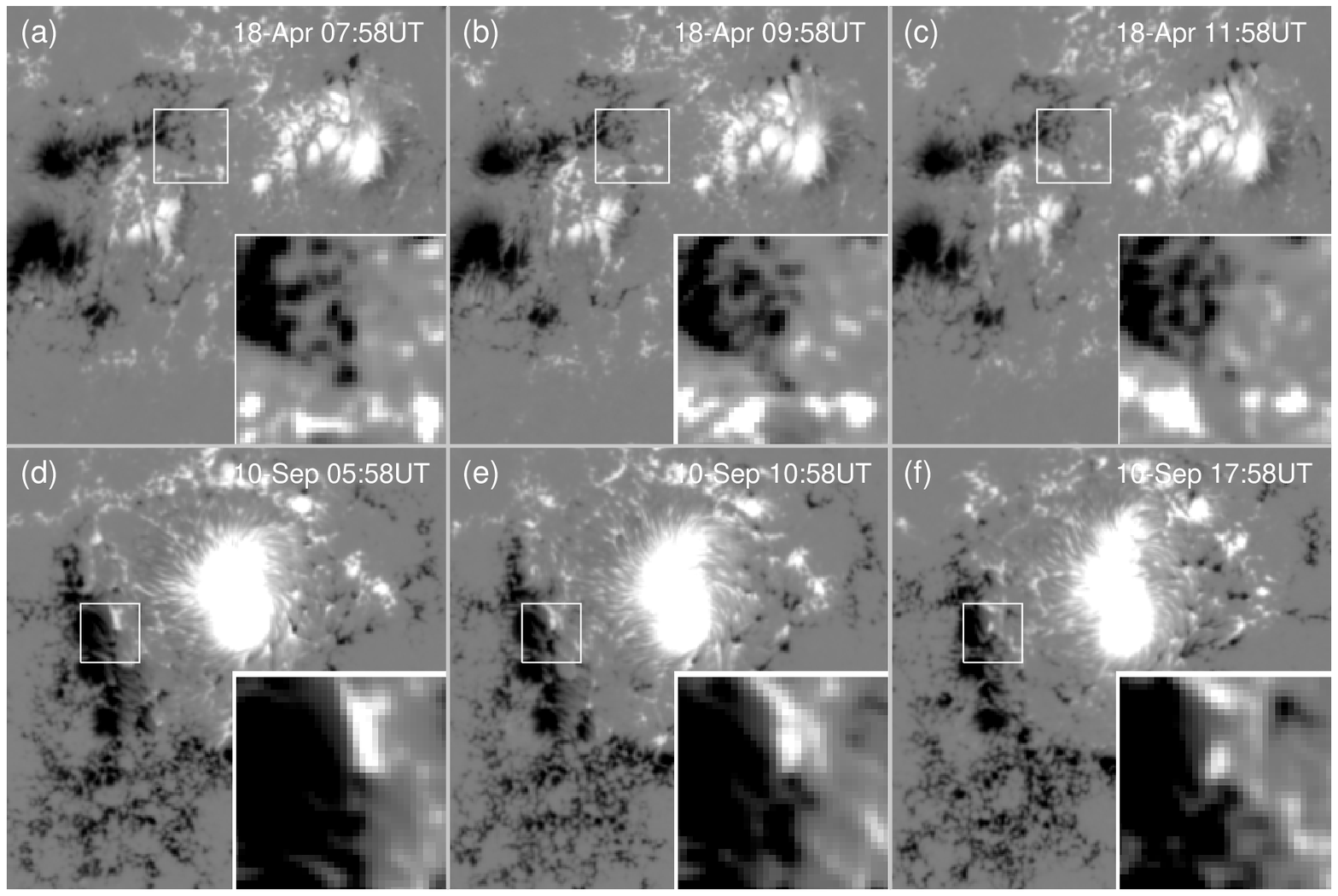}}
\caption{(a--c) A sequence of HMI line-of-sight magnetograms showing the convergence flow and magnetic cancellation near the formation site (white boxes) of the MFR1. (d--f) HMI line-of-sight magnetograms displaying the shearing flow and cancellation near the formation site (white boxes) of the MFR2.}
(Animations of this figure are available in the online journal.)
\label{fmag}
\end{figure*}

\begin{figure*}
\vspace{-0.6\textwidth}
\center {\includegraphics[width=14cm]{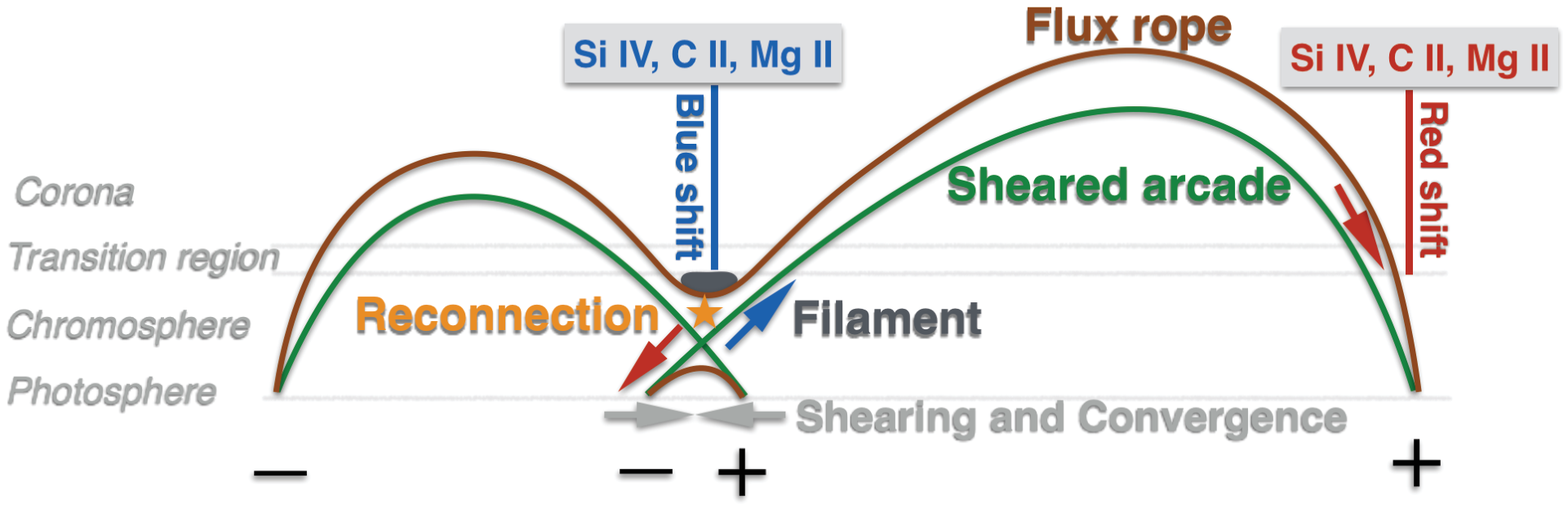}}
\vspace{-0.4\textwidth}
\caption{Schematic drawing of the formation of the MFR (brown) through the reconnection (yellow star) between two sheared arcades (green) in the chromosphere driven by the shearing and converging flows in the photosphere (grey arrows). The filament material (grey) is suspended in the magnetic dip of the MFR. The blueshift (blue arrow) and the redshift (red arrow) of the spectral lines (Si IV, C II, and Mg II) are observed at the center and footpoints of the MFR, respectively.}
\label{fcartoon}
\end{figure*}

\section{Summary and Discussions}
In this paper, we investigate the formation and initiation of two MFRs through imaging and spectral observations by AIA and \textit{IRIS}. The EUV observations show that the magnetic field involved in the formation of the MFRs is primarily within in the core field of the active regions and is organized as two groups of sheared arcades. Their appearance only in the 131 {\AA} and 94 {\AA} passbands but not in other cooler passbands indicates that the arcades have a temperature of $\ge$7 MK. Through examining a sequence of line-of-sight magnetograms of the HMI (Figure \ref{fmag} and attached movies), we find that the converging and shearing flows, as well as magnetic cancellation, occur near the main polarity inversion line. For the MFR1, some dispersive and weak opposite polarities move ceaselessly toward each other (Figure \ref{fmag}a--\ref{fmag}c), resulting in magnetic cancellation near the possible formation site of the MFR1. For the MFR2, besides magnetic cancellation driven by the shearing flow, an obvious rotation of the preceding positive polarity also efficiently shears the arcades (Figure \ref{fmag}d--\ref{fmag}f). Based on the observational properties, we suggest that the reconnection driven by the photospheric flows most likely take place and play a vital role in the formation of MFRs. On one hand, the reconnection may connect the head and end of two sheared arcades to form a continuous sigmoidal loop and accumulate the poloidal field of the MFR. On the other hand, it simultaneously heats the plasma in the MFR to make it only visible in the high temperature passbands of the AIA. 

\textit{IRIS} SJI images also provide valuable information to understand the formation of the MFRs. Except for the brightening at the footpoints, other parts of the MFR1 are invisible in all the SJI images and AIA UV images, implying that its formation probably occurs mainly in the corona. While for the MFR2, it shows up not only in all AIA passbands but also in the \textit{IRIS} 1400 {\AA} and 2796 {\AA} passbands, indicating that its formation likely takes place in the transition region and/or the chromosphere. Moreover, the finding of the cospatiality of the  cool filamentary materials, which are visible in the AIA 304 {\AA} and 171 {\AA} passbands, with the brightened sigmoidal field also supports the hypothesis that the formation of the MFR2 may extend to the chromosphere. 

The different formation altitudes between the MFR1 and MFR2 emphasizes the diversity of the reconnection locations, which can range from the photosphere to the corona. The results are consistent with the recent conclusions by \cite{cheng14_formation}, in which they studied the long-term evolution of a sigmoidal active region and found that the twist in the core field gradually increases with reconnection continuously converting the sheared arcades to the sigmoidal field both at the bald-patch in the photosphere and in the hyperbolic flux tube in the corona.

\textit{IRIS} spectral data further disclose some new features of the formation and initiation of the MFRs. For the MFR1, the Si IV, C II, and Mg II lines present a redshift and slightly large non-thermal velocity at the footpoints during the initiation phase. We interpret this as being caused by reconnect-induced outflow, which propagates along the legs of the MFR1 and finally reaches the lower atmosphere at the footpoints. Note that in flare ribbons, one can observe stronger redshift (or blueshift) and broadening of various lines that are mostly caused by chromospheric evaporation and/or condensation driven by the non-thermal electrons generated during the fast reconnection \citep{fisher85,liying11,tian14_reconnection}. In the case of the initiation of the MFR1, however, the redshift and broadening at the footpoints may instead be related to the outflow driven by the reconnection surrounding or inside the MFR1 that seems more gentle. Recent detection of a pre-flare enhancement of the non-thermal velocity at the footpoints of a CME \citep{harra13} also supports such an interpretation.

For the MFR2, the \textit{IRIS} slit is positioned at its formation site. Generally, the Si IV, C II, and Mg II lines show an extremely strong blueshift and non-thermal broadening at the most brightened pixels in the formation site. If it is still explained in terms of the reconnection scenario, then it requires that the reconnection has to occur in a place low enough so that either upward moving energetic electrons or reconnection outflow can heat the plasma above, and thus result in the blueshift and broadening of the Mg II, C II, and Si IV lines. Nevertheless, at some weakly brightened pixels at the formation site, the three lines display complicated patterns that seem puzzling. The Mg II line is blueshifted without ambiguity. However, the Si IV and C II lines exhibit a stronger red wing. Checking the line profiles, we are more inclined to the interpretation of an absorption at the blue wing, which can also be related to an upward moving plasma with a temperature not high enough to make it emissive relative to the Si IV and C II background emission \citep[also see;][]{rouppe14}. The reconnection scenario also expects the appearance of a downflow \citep[e.g.,][]{innes97,peter14}. However, for the MFR2, we only detect the upflow. A possible reason is that the downflow is kinetically less obvious than the upflow and could be smeared out in the spectra owing to the radiative transfer effect along the line of sight.

As a conclusion, based on the AIA and \textit{IRIS} joint observations, we propose a cartoon for the formation of two MFRs, as shown in Figure \ref{fcartoon}. Two sheared arcades approach each other driven by the shearing and converging flows. The head of the left arcade and the end of the right arcade reconnects near the main polarity inversion line, forming a smaller loop and a longer MFR. Due to magnetic tension, the submergence of the small loop near the polarity inversion line, manifested as magnetic cancellation, is expected \citep{martin98}. Meanwhile, the twisted MFR may rise up to the corona or still stay in the chromosphere depending on if there are filament materials suspended in the dips of the MFR. In the traditional tether-cutting model \citep{moore01}, the location of reconnection is not specified. Here, we suppose that the reconnection site is in the middle chromosphere, where the non-thermal particles accelerated during the reconnection can move upwards and lead to the large blueshift and strong broadening of the Si IV, C II, and even Mg II lines. The reconnection can also produces an obvious upward outflow, which can propagate along the field line of the MFR to the footpoint, thus making the Si IV, C II, and Mg II lines redshifted and weakly broadened.

\acknowledgements We are grateful to the referee for his/her constructive comments that helped improve the manuscript. We also thank Hui Tian and Ying Li for the discussions on data calibration. \textit{SDO} is a mission of NASAs Living With a Star Program. \textit{IRIS} is a NASA small explorer mission developed and operated by LMSAL with mission operations executed at NASA Ames Research center and major contributions to downlink communications funded by the Norwegian Space Center (NSC, Norway) through an ESA PRODEX contract. X.C., M.D.D., and F.C. are supported by NSFC under grants 11303016, 11373023, and NKBRSF under grants 2011CB811402 and 2014CB744203. X.C. is also supported by Key Laboratory of Solar Activity of National Astronomical Observatories of the Chinese Academy of Sciences by Grant KLSA201311.

\end{document}